\documentclass[reprint, amsmath,amssymb, aps, superscriptaddress
]{revtex4-1}
\usepackage{amssymb}
\usepackage{amsmath}
\usepackage{epsfig}
\usepackage{color}
\usepackage{graphics, graphicx}
\usepackage{bbold}
\usepackage{psfrag}
\usepackage{mathcomp}
\usepackage{subfigure}
\usepackage{verbatim}
\usepackage{color}
\usepackage{float}
\usepackage[colorlinks, citecolor=blue]{hyperref}

\graphicspath{{img/}}

\begin{document}
\preprint{APS/123-QED}
\title{Highly polarized one-dimensional Fermi gases near a narrow $p-$wave resonance}
\author{Yinfeng Ma}
\affiliation{Beijing National Laboratory for Condensed Matter Physics, Institute of Physics, Chinese Academy of Sciences, Beijing 100190, China}
\affiliation{University of Chinese Academy of Sciences, Beijing 100049, China}
\author{Xiaoling Cui}
\affiliation{Beijing National Laboratory for Condensed Matter Physics, Institute of Physics, Chinese Academy of Sciences, Beijing 100190, China}
\affiliation{Songshan Lake Materials Laboratory, Dongguan, Guangdong 523808, China}
\date{\today}

\begin{abstract}
 Based on the recently developed interaction renormalization for the one-dimensional $p$-wave interaction, we study the problem of a single impurity immersed in a highly polarized Fermi sea. They interact through a narrow $p$-wave Feshbach resonance, so the effective range $r_{0}$ naturally appears in the system. We use the variational approach limited to single-particle-hole excitations on top of the unperturbed Fermi sea. The polaron exhibits two branches of solutions, namely, the attractive and repulsive branches, with varying scattering length across the resonance. We calculate the energy spectrum, residue and effective mass for each of the branches. We compare the polaronic energy with the energy of the dressed molecule, and find that the molecular state is energetically favored when increasing the interaction strength. The critical interaction strength for the polaron-to-molecule transition will shift to the BCS side of the $p$-wave resonance as the effective range increases.

\end{abstract}
\pacs{Valid PACS appear here}
\maketitle
\section{INTRODUCTION}
\label{introdution}

Highly polarized atomic gas, which consists of a single impurity immersed in a bath of majority atoms, gives rise to the quasiparticle excitation known as the polaron \cite{Chevy2010,Massignan2014}. According to the statistics of majority atoms, the polaron system can be classified as Bose and Fermi polarons. In recent years, there have been extensive investigations into these two types of polarons, which have also been successfully realized in cold atomic gases experiments \cite{PhysRevLett.102.230402,PhysRevLett.103.170402,Kohstall2012,Koschorreck2012,PhysRevLett.108.235302,Scazza,Valtolina2017,Hu2016a,PhysRevLett.117.055302}. The interaction strength can be tuned over a wide range using Feshbach resonances, and the experiments have detected both the attractive and repulsive branches of polarons via radio-frequency spectroscopy \cite{PhysRevLett.102.230402,Kohstall2012,PhysRevLett.117.055302}. In particular, for three-dimensional (3D) and 2D Fermi polaron systems in the attractive branch, it has been demonstrated that polaron-to-molecule transitions do exist by using the diagrammatic Monte Carlo approach and the  variational wave function in the ground state \cite{Prokofev2008,Prokofev2008a,Combescot2009a,Mora2009a,Punk2009, Bruun2010, Mathy2010, Schmidt2011c, Parish2011}, which has also been successfully detected in the 3D highly polarized Fermi gas \cite{PhysRevLett.102.230402,Kohstall2012}. In comparison, the 1D Fermi polaron does not show any sharp transition with increasing interaction strength \cite{McGuire1966,Giraud2009a,Parish2011,PhysRevA.94.043645}, which is distinct from 3D and 2D systems.

So far, most studies on polaron properties are focused on $s$-wave interacting systems. Nevertheless, discussions about $p$-wave polarons are rate, except for a few theoretical works \cite{Levinsena}. It has been shown that the $p$-wave polaron in three dimensions displays distinct features as compared to the 3D $s$-wave polarons \cite{Levinsena}. It is natural to ask how the polaron picture behaves in low dimension such as in one dimension. Recently, the interaction renormalization of effective two-body scattering has been established in the 1D $p$-wave system \cite{Cui2016}, which paves the way for further investigation of the polaron physics in this system.

In this work, we investigate the polaron physics for 1D atomic Fermi gas across a $p$-wave resonance. We have adopted the variational wave function up to the single-particle-hole excitation, which has been extensively used to treat polaron problems in $s$-wave interacting systems \cite{Chevy2006,Combescot2008,Cui2010,Parish2011,Massignan2011,PhysRevA.83.021603,Ngampruetikorn2012}. We have used the two-channel model to describe the $p$-wave system, which naturally incorporates the effective range effect near a narrow $p$-wave resonance. Using the recently developed interaction renormalization for the 1D $p$-wave system, assisted by the variational wave function approach, we have studied the energy spectrum, residue, and effective mass of the Fermi polaron. It is found that the polaron exhibits two branches of solutions, namely, the attractive and repulsive branches, with varying scattering length across the resonance. We plot the energy spectrum, residue and effective mass for each of the branches, and compare the polaron energy with the energy of the dressed molecule, whose wave function is limited to single-particle-hole excitation. We find that the attractive polaron branch becomes energetically unstable owing to the molecule formation with increasing interaction strength, signifying the polaron-molecule transition. This is distinct from the $s$-wave case in one dimension,  where such a transition is absent\cite{McGuire1966,Giraud2009a,Parish2011,PhysRevA.94.043645}. We also show that the critical interaction strength for the polaron-molecule transition shifts to the BCS side of the $p$-wave resonance as the effective range increases. These results may be detected in future experiments on p-wave Fermi gas confined in quasi-1D geometry.

The structure of this paper is as follows. In Sec. \ref{twoch}, we discuss the narrow  Feshbach resonance in a two-channel model. In Sec. \ref{tail}, we study polaronic and molecular states using the variational approach, as well as polaron-molecule transitions. We summarize in Sec. \ref{conclusion}.

\section{Two-channel model}
\label{twoch}
\begin{figure}[H]
 \centering
 \includegraphics[width=0.45\textwidth]{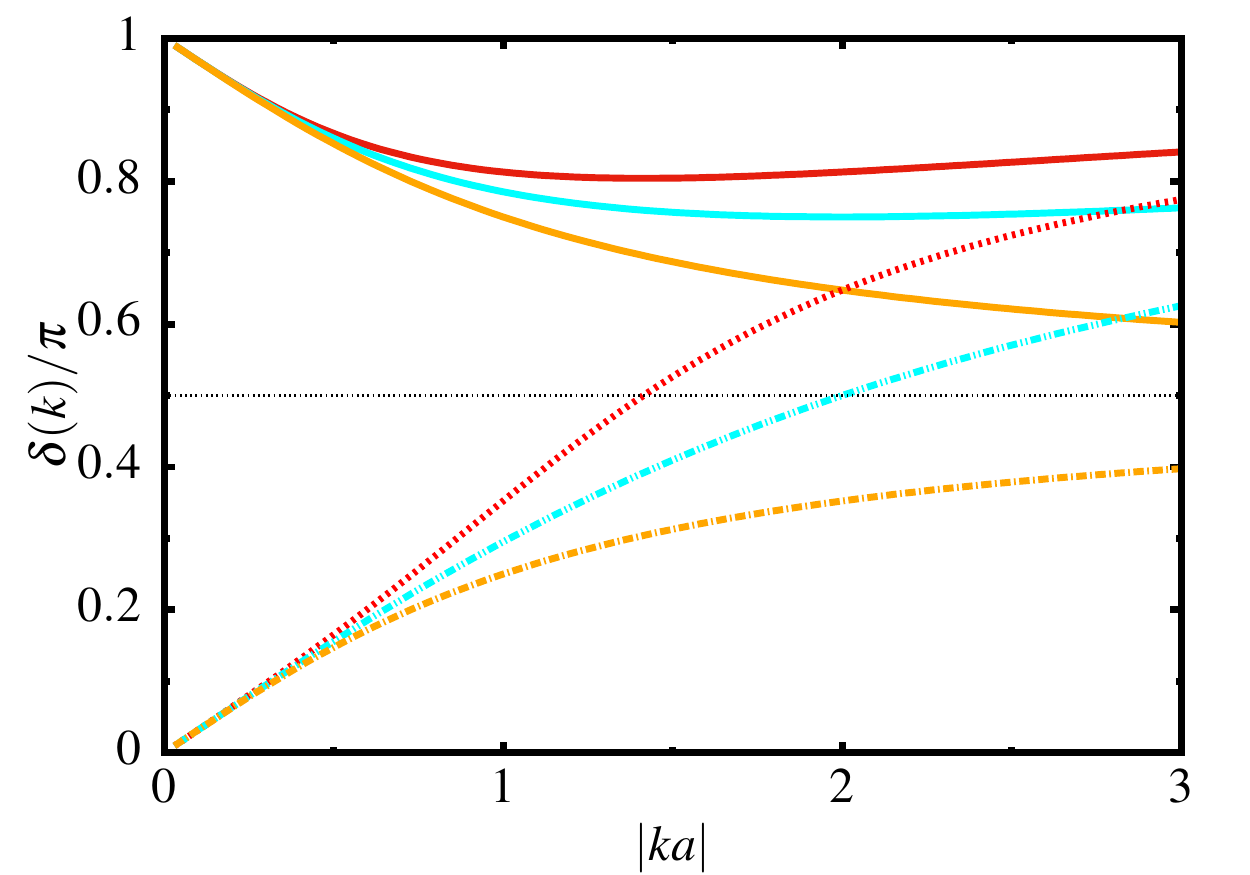}
 \caption{Phase shifts with positive (solid curves) and negative (dashed curves)scattering length $a$. The effective range is chosen as $r_0/|a|=-1,-0.5,0$ from top to bottom for both curves. The horizontal dotted line corresponds to the unitarity limit.}
\label{phases}
\end{figure}

We describe a two-component Fermi system in one dimension. The system is composed of one spin-$\downarrow$ and $N $ spin-$\uparrow$ atoms ,d , both with equal mass $m$. We assume that there is no interaction among the spin-up atoms, and that the single spin-down atom interacts with the spin-up fermion resonantly via a $p$-wave narrow Feshbach resonance. We use the two-channel model to describe this resonant interaction: The particles exist either in the form of atoms in the so-called open channel, or in the form of a tightly bound molecule, in the closed channel. The second-quantized Hamiltonian for the system can be written:
\begin{equation}\label{hp}
\begin{split}
 H=&\sum_{p\sigma}\epsilon_p c_{p\sigma}^\dagger c_{p\sigma}+\sum_p\left(\frac{\epsilon_p}{2}+\nu\right)b_p^\dagger b_p\\
 &+\frac{g}{\sqrt L}\sum_{pq}q\left(b_p^\dagger c_{p/2+q\uparrow}c_{p/2-q\downarrow}+H.c.\right).
 \end{split}
 \end{equation}
 Here $c_{p\sigma}^\dagger$ is the creation operators of an atom with momentum $p$, spin $\sigma$ ($\sigma=\uparrow,\downarrow$) and kinetic energy $\epsilon_p=p^2/2m$. $b_p^\dagger$ denotes the bosonic creation operator of a molecule with momentum $p$, and mass $2m$. The $c_{p\sigma}$ obey the usual fermionic anticommutation relation, $\{c_{p\sigma},c_{k\mu}^\dagger\}=\delta_{pk}\delta_{\sigma\mu}$. The molecule obeys Bose statistics, so $[b_{p},b_{k}^\dagger]=\delta_{pk}$. In additipn, $L$ is the length of the system, and $\nu$ is the molecular bare detuning energy. The last term describes the coupling between the closed-channel molecule and atoms in the open channel, with the momentum-dependent coupling constant $gq$, owing to the $p$-wave interaction. In the ladder approximation, the two-body  $T$ matrix in the present problem is obtained as \cite{Cui2016}
 \begin{equation}\label{t}
\begin{split}
T_{kk^\prime}=kk^\prime\left(\frac{ L(E-\nu)}{g^2}+\sum_q\frac{q^2}{2\epsilon_q}+\frac{imkL}{2}\right)^{-1}.
\end{split}
\end{equation}
In order to relate the parameter of the model to the physical parameters, we compare with the low energy expansion of the 1D $p$-wave scattering amplitude $f(k,k^\prime)=-\frac{mL}{2}T_{kk^\prime}$:
\begin{equation}\label{f}
f(k,k^\prime)=\frac{kk^\prime}{-a^{-1}+1/2r_0k^2-ik}
\end{equation}
and the low-energy phase shift
\begin{equation}\label{phase}
k\cot\delta(k)=-a^{-1}+1/2r_0k^2
\end{equation}
Here $a$ is the scattering length and $r_0$ is the $p$-wave effective range. Figure.\ref{phases} shows the phase shift $\delta(k)$ at various scattering lengths and effective ranges. While in the
low-momentum region $\delta(k)$ does not depend on $r_0$, in the high-momentum region, it depends on $r_0$.

Combining Eq. \ref{t} and Eq. \ref{f}, the bare values of the detuning $\nu$ and the Feshbach coupling strength $g$ can be related to the physical $p$-wave scattering length $a$ and the interaction range $r_0$ via
\begin{equation}
\begin{split}
\frac{m}{2a}&=-\frac{\nu}{g^2}+\frac{1}{L}\sum_q\frac{q^2}{2\epsilon_q}\\
r_0&=-\frac{4}{m^2g^2}
\end{split}
\end{equation}
Note that the parameter $r_0$ is naturally negative. The poles of the scattering amplitude determine the energy of the bound states of the interaction potential. We see that Eq. \ref{f} has a single pole, describing a dimer with binding energy
\begin{equation}
E_{d}=-\frac{1}{ma_\star^2}
\end{equation}
where $a_\star=2R/\left(\sqrt{1+4R/a}-1\right)$, with $R=-r_0/2$. For $R\ll a$, we recover the universal energy $E_d=-1/ma^2$, whereas we get $E_d=-1/mRa$ when $R\gg a$.
\section{polaronic and molecular variational ansatz}
\label{tail}
\subsection{polaronic state}
We explore the system on a narrow Feshbach resonance using a simple variational ansatz with single-particle-hole excitations of the Fermi sea. We consider the general case of a non-zero total momentum for the system:
\begin{equation}
\label{wfp}
\begin{split}
	|{\cal P}(p)\rangle=&\alpha_0 c_{p\downarrow}^\dagger |FS^{N}_\uparrow\rangle+{\sum_{q}}^\prime\beta_{q}b_{p+q}^\dagger c_{q\uparrow}|FS^{N}_\uparrow\rangle\\
	&+{\sum_{kq}}'\alpha_{k q}c_{p+q-k\downarrow}^\dagger c_{k\uparrow}^\dagger  c_{q\uparrow}|FS^{N}_\uparrow\rangle
  \end{split}
\end{equation}
where $p$ indicates the center-of-mass momentum of the polaron, $|FS^{N}_\uparrow\rangle$ is the N-particle Fermi sea, and $\epsilon_F=k_F^2/2m$. Here and in the following, the prime to the summation means that the sums are restricted to $|q|<k_F$ and $|k|>k_F$, respectively, where $k_F$ is the Fermi momentum of underlying Fermi sea.

The ground-state energy can be obtained by calculating the expectation value $\langle{\cal P}(p)|H-E|{\cal P}(p)\rangle$, taking the derivatives with respect to the variational parameters $\alpha_0,\beta_{q}$, and $\alpha_{kq}$, and setting them equal to zero. This leads to the set of coupled equations

\begin{equation}
\begin{split}
	\left(E-\epsilon_p\right)\alpha_0&=\frac{g}{\sqrt L}{\sum_q}^\prime\frac{q-p}{2}\beta_{q}\\
\left(E-\frac{\epsilon_{p+q}}{2}-\nu+\epsilon_q\right)\beta_{q}&=\frac{g}{\sqrt L}\frac{q-p}{2}\alpha_0\\
	&\quad +\frac{g}{\sqrt L}{\sum_{k}}^\prime\frac{2k-p-q}{2}\alpha_{kq}\\
\left(E-\epsilon_{p+q-k}-\epsilon_k+\epsilon_q\right)\alpha_{kq}&=\frac{g}{\sqrt L}\frac{2k-p-q}{2}\beta_{q}
\end{split}
\end{equation}
where the ground-state energy $E$ is measured with respect to the $N$-particle Fermi sea. After a straightforward calculation, this yields a self-consistent equation for the ground-state energy of the polaron:
\begin{equation}
\begin{split}
	E=\epsilon_p&+{\sum_{q}}'\frac{(q-p)^2}{4L}\left[\frac{m^2|r_0|}{4}\left(E-\epsilon_{p+q}/2+\epsilon_q\right)\right.\\
	  &\left.+\frac{m}{2a}-\frac{1}{L}\sum_{k}\frac{k^2}{2\epsilon_k}-{\sum_{k}}'\cfrac{(2k-p-q)^2}{4LE_{kq}}\right]^{-1}
\end{split}
\label{polen2}
\end{equation}
where
\begin{equation}
\begin{split}
E_{kq}&=E-\epsilon_{p+q-k}-\epsilon_k+\epsilon_q
\end{split}
\end{equation}

\begin{figure}[H]
 \centering
 \subfigure[]{\includegraphics[width=0.45\textwidth]{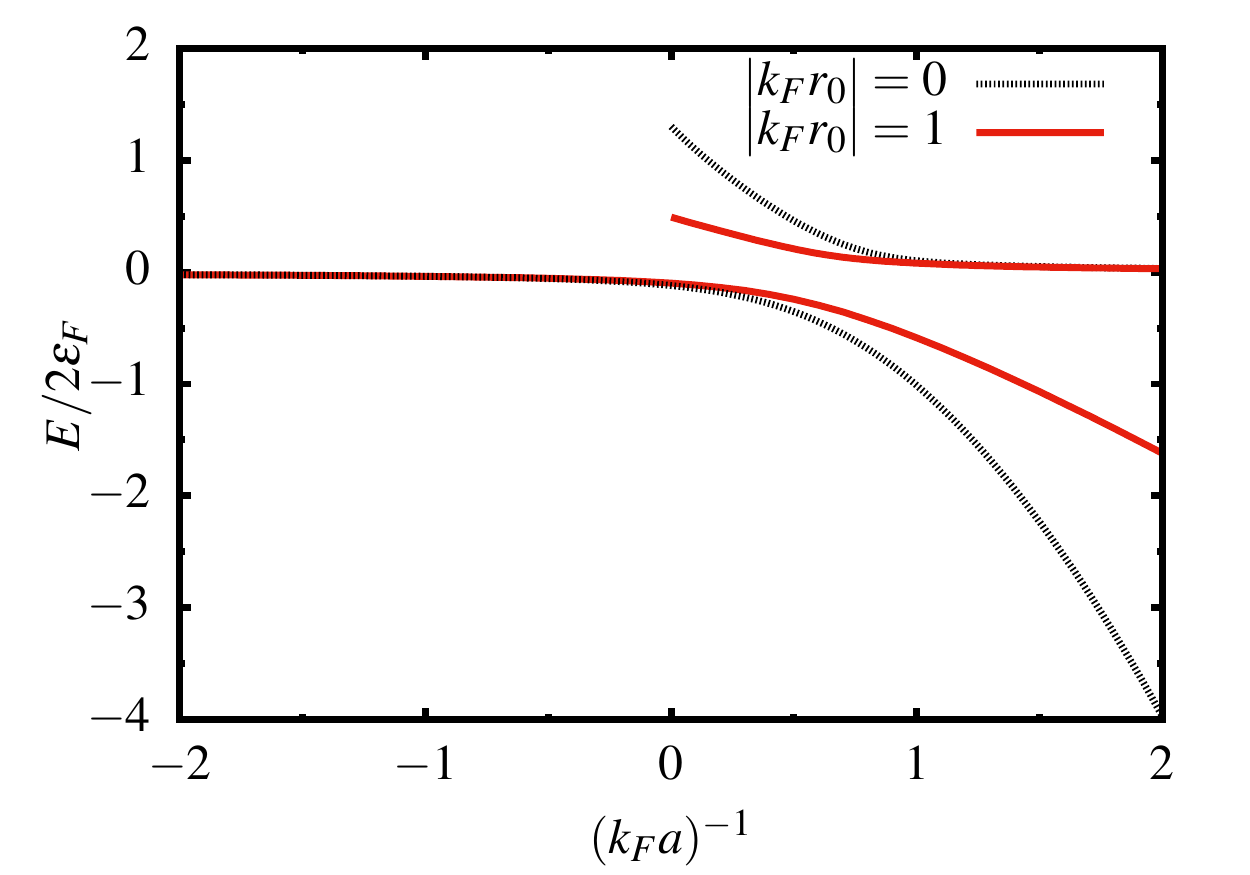}\label{ena}}
 \subfigure[]{\includegraphics[width=0.45\textwidth]{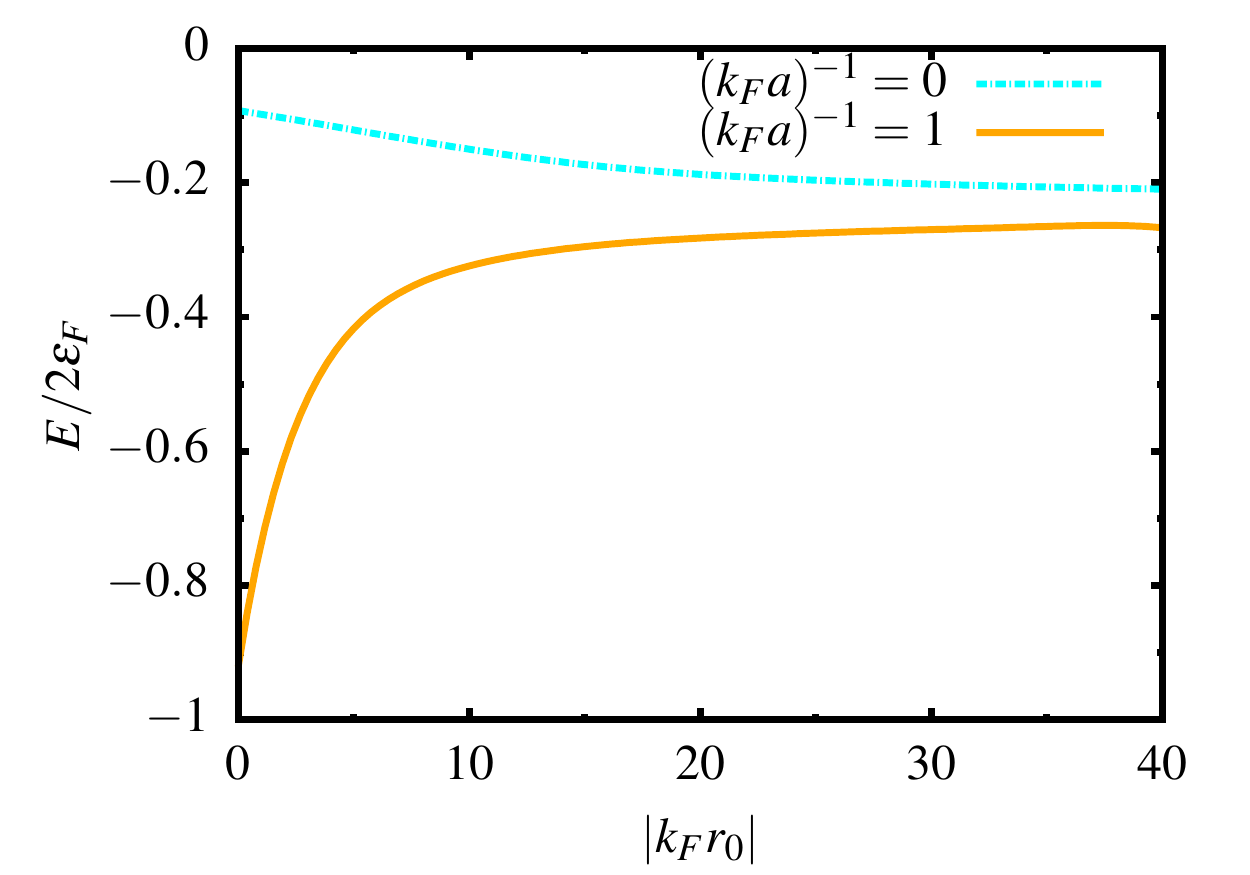}\label{enr}}\\
 \caption{Polaron energy $E$ in units of the $2\epsilon_F$. \subref{ena} Two branches with different effective ranges versus $(k_Fa)^{-1}$. $|k_Fr_0|=0$ (black dashed line); $|k_Fr_0|=1$ (red solid line).\subref{enr} Lower branch of the polaron energy with different interaction strengths as a function of $|k_Fr_0|$: $(k_Fa)^{-1}=0$ (cyan dot-dashed line), and $(k_Fa)^{-1}=1$ (orange solid line).}
\label{pen}
\end{figure}

We numerically solve Eq. \ref{polen2} with $p=0$ for various $|k_Fr_0|$, yielding the polaron energy presented in Fig. \ref{pen}. As shown in Fig. \ref{ena}, the polaron exhibits two branches of solutions $E_\pm$, namely, the attractive and repulsive branches, with varying scattering length across the resonance. For a given $|k_F r_0|$, the attractive and repulsive polaron energies monotonically decreases with increasing $(k_Fa)^{-1}$. In Fig. \ref{enr} we display how the attractive polaron energy for two different interaction strengths is affected by effective range $|k_Fr_0|$. The attractive polaron energy will increase at $(k_Fa)^{-1}=1$ as $|k_Fr_0|$ increases. Contrary to the three-dimensional case \cite{Qi2012a,Massignan2012a,Trefzger2012a}, at $(k_Fa)^{-1}=0$, the energy will decrease. In the limit $|k_Fr_0|\to\infty$, the attractive polaron energy saturates to $-0.5\epsilon_F$.

In general, all polaron properties are given in terms of the self-energy $\Sigma(p,E)$ of the spin-down atom. Using the ladder approximation at $T=0$, as pointed out for $s$-wave polaron system in \cite{Combescot2007}, we can obtain the self-energy for 1D $p$-wave polarons
\begin{equation}
  \begin{split}
	  \Sigma(p,E)&={\sum_{q}}'\frac{(q-p)^2}{4L}\left[\frac{m^2|r_0|}{4}\left(E-\epsilon_{p+q}/2+\epsilon_q\right)\right.\\
	  &\left.+\frac{m}{2a}-\frac{1}{L}\sum_{k}\frac{k^2}{2\epsilon_k}-{\sum_{k}}'\cfrac{(2k-p-q)^2}{4LE_{kq}}\right]^{-1}
    \end{split}
\label{selfenergy}
\end{equation}
The energies of the two polaron branches are given  by the two solutions of the equation
\begin{equation}
\label{eq:1}
E=\epsilon_{p\downarrow}+Re[\Sigma(p,E)]
\end{equation}
where $ Re$ extracts the real part. In addition to the energy, the polarons are also described by the residue and effective mass. The polarons residues and effective masses at $p=0$ can be obtained as:
\begin{equation}
Z_{\pm}=\left(1-\frac{\partial}{\partial E}Re\Sigma(0,E_{\pm})\right)^{-1}
\end{equation}
and
\begin{equation}
\frac{m_{\pm}^\star}{m}=\frac{1}{Z_{\pm}}\left(1+\frac{\partial }{\partial \epsilon_{p\downarrow}}Re \Sigma(0,E_{\pm})\right)^{-1}
\end{equation}
\begin{figure}[H]
\includegraphics[width=0.45\textwidth]{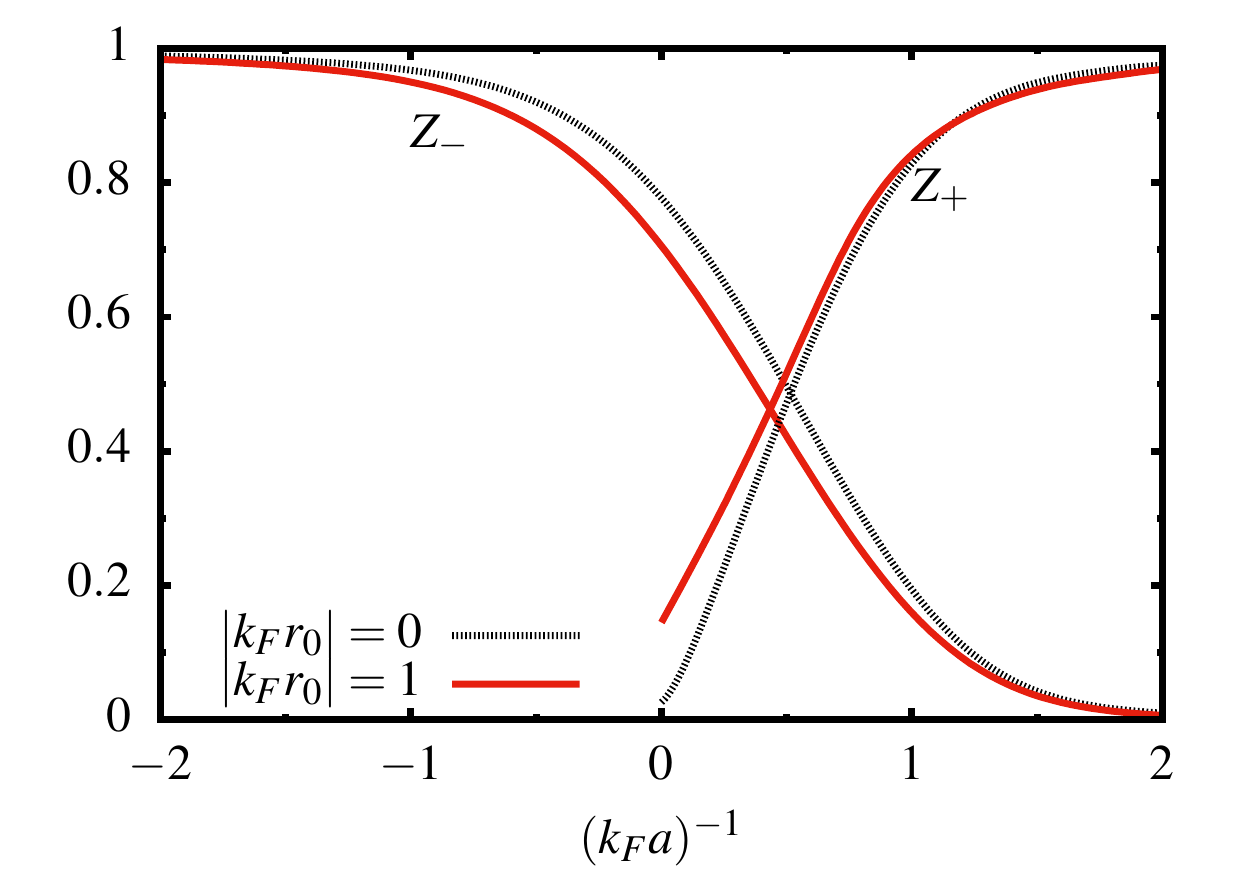}
\caption{Quasiparticle residue $Z$ of the two polaron branches as a function of $(k_{F}a)^{-1}$ for two different $|k_Fr_0|$: $|k_Fr_0|$=1(Red solid lines) and $|k_Fr_0|=0$(black dashed lines).}
\label{fig:z}
\end{figure}
In Fig. \ref{fig:z} we plot the quasiparticle residue for $|k_Fr_0|=0$ and $1$, where $Z_+, and Z_-$ are the repulsive and attractive polaron residues, respectively. We find that $Z_-$ monotonically decreases with increasing $(k_Fa)^{-1}$, while $Z_+$ behave oppositely. In the BCS limit $a\to0^-$, we have $Z_-\to 1$, and in the Bose-Einstein-Condensation (BEC) limit $a\to0^+$, $Z_+\to 1$. Similar to 3D $s$-wave systems $Z_-$ remains a finite value, even at large interaction strengths.
\begin{figure}[H]
\includegraphics[width=0.5\textwidth]{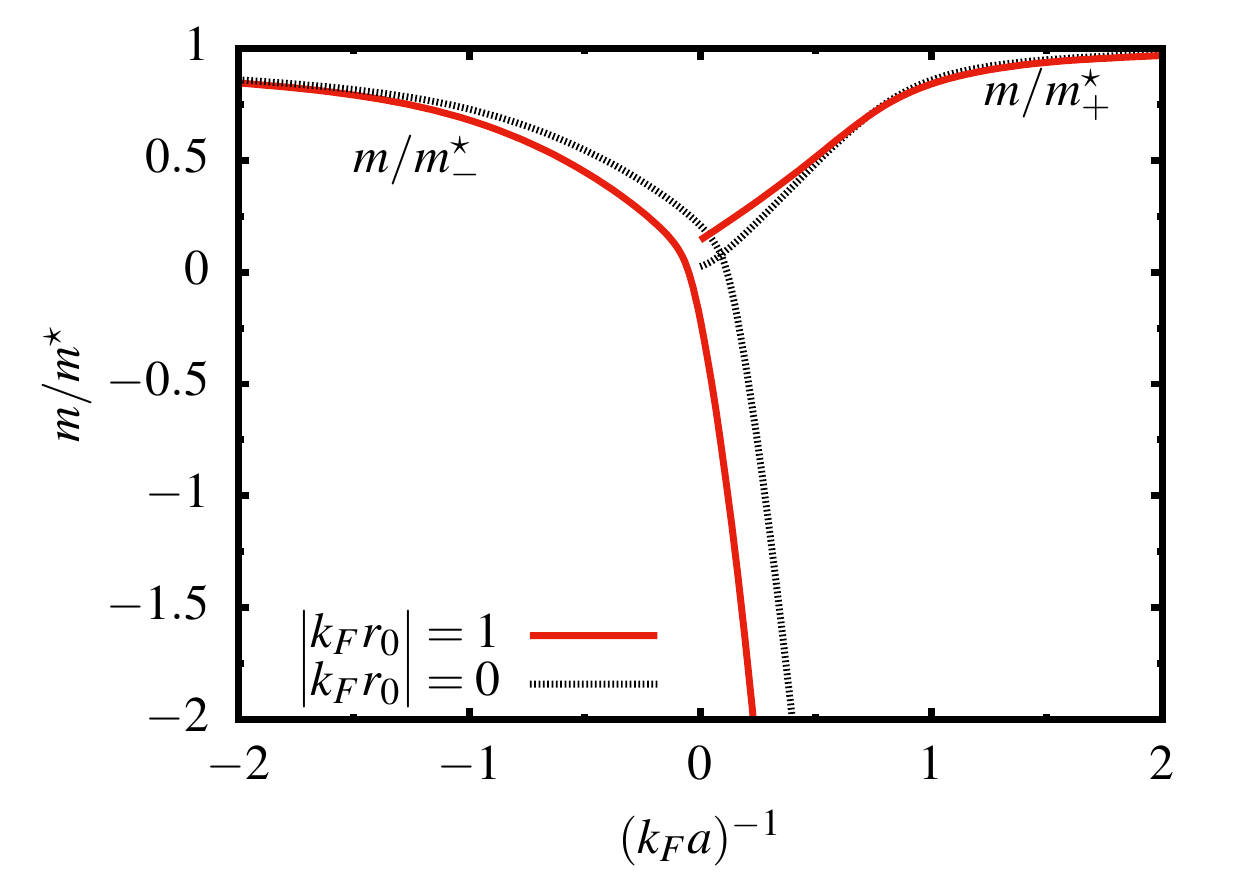}
\caption{Inverse effective mass $1/m^\star$ of the polaron , for different value of the effective range , $|k_Fr_0|$=1(red solid lines), and $|k_Fr_0|=0$ (black dashed lines).}
\label{fig:mass}
\end{figure}
At small momenta, the dispersion can be written as
\begin{equation}
E(p)=E(0)+\frac{p^2}{2m^\star}
\label{quasiparticle}
\end{equation}
The effective mass $m^\star$ of the polaron for two different effective range is shown in Fig.\ref{fig:mass}. Naturally, we see that in the BEC limit $m_+^\star$ approaches the bare impurity mass $m$ and the effect of $|k_Fr0|$ can be ignored.  $m_+^\star$ rises dramatically in the unitary regime, because the polaron gets increasingly dressed by particle-hole pairs. As the particle-hole fluctuations become suppressed as the effective range increases, the mass of the repulsive polaron shows tendency to slowly increase in the unitary regime. For finite $r_0$ , as the attractive polaron goes beyond a critical strengths $(k_Fa)^{-1}$ , the effective mass diverges and becomes negative. This is the precursor of the polaron-molecule transition as studied below.

\subsection{Dress molecular state and polaron-molecule transition}
For the molecular state, we use the variational wave function which includes one particle-hole contributions with total momentum $p$, analogous to the  3D case \cite{Punk2009}:
\begin{equation}\label{wfm}
\begin{split}
	|{\cal M}(p)\rangle&=\left(\beta_0 b_p^\dagger+{\sum_{kq}}'\beta_{kq}b_{p+q-k}^\dagger c_{k\uparrow}^\dagger c_{q\uparrow}+{\sum_k}'\alpha_k c_{p-k\downarrow}^\dagger c_{k\uparrow}^\dagger\right.\\
	&\left.+{\sum_{kk^\prime q}}'\alpha_{k^\prime k q}c_{p+q-k-k^\prime\downarrow}^\dagger c_{k^\prime\uparrow}^\dagger c_{k\uparrow}^\dagger  c_{q\uparrow}\right)|FS^{N-1}_\uparrow\rangle
  \end{split}
\end{equation}

where $|FS^{N-1}_\uparrow\rangle$ corresponds to removeing one spin-up from Fermi surface of $N$ fermions from the Fermi surface of N fermions. The minimization procedure yields the set of coupled equations for the energy of the molecule
\begin{equation}
\begin{split}
	&\left(E+\epsilon_F-\frac{\epsilon_p}{2}-\nu\right)\beta_0=\frac{g}{\sqrt L}{\sum_k}'(k-\frac{p}{2})\alpha_k\\
&\left(E+\epsilon_F-\epsilon_{p-k}-\epsilon_k\right)\alpha_k=\frac{g}{\sqrt L}(k-\frac{p}{2})\beta_0\\
	&-\frac{g}{\sqrt L}{\sum_q}'\frac{q-p+k}{2}\beta_{kq}\\
&\left(E+\epsilon_F-\frac{\epsilon_{p+q-k}}{2}-\nu-\epsilon_k+\epsilon_q\right)\beta_{kq}=\\
	&-\frac{g}{\sqrt L}\frac{q-p+k}{2}\alpha_k+\frac{2g}{\sqrt L}{\sum_{k^\prime}}'\frac{2k^\prime+k-p-q}{2}\alpha_{k^\prime kq}\\
&\left(E+\epsilon_F-\epsilon_{p+q-k-k^\prime}-\epsilon_k-\epsilon_{k^\prime}+\epsilon_q\right)\alpha_{k^\prime kq}=\\
&\frac{g}{2\sqrt L}\left(\frac{2k^\prime +k-p-q}{2}\beta_{kq}-\frac{2k+k^\prime-p-q}{2}\beta_{k^\prime q}\right)
\end{split}
\label{molen}
\end{equation}

where the ground-state energy $E$ is measured with respect to the $N$-particle Fermi sea.
\subsubsection{No particle-hole excitation}
We first neglect the contribution of particle-hole excitation in Eq. \ref{wfm}, i.e., setting $\alpha_{k^\prime kq}=0$ and $\beta_{kq}=0$. It is the lowest-order variational wave function for the molecule. Minimizing the expectation value $\langle{\cal M}(p)|H-E|{\cal M}(p)\rangle$ yields an implicit equation for the molecular energy:
\begin{equation}
\label{nophm}
\left(E+\epsilon_F-\frac{\epsilon_p}{2}-\nu\right)=\frac{g^2}{L}\sum_k\frac{(k-\frac{p}{2})^2}{E+\epsilon_F-\epsilon_{p-k}-\epsilon_k}
\end{equation}
when $p=0$ and $x=E/2\epsilon_F+0.5$ , Equation \ref{nophm} reduces to
\begin{equation}
\label{nphm}
\frac{\pi}{2k_Fa}-1+\frac{\pi}{4}|k_Fr_0|x=\sqrt{-x}\arctan{\sqrt{-x}}
\end{equation}
The bare molecular state energy is shown in Fig. \ref{mpen} with a green dot-dashed line. As the interaction strengths increase, the system favors the molecular state over the polaronic state. In the BEC limt $a\to0^+$, Eq. \ref{nphm} gives rise to a ground-state energy
\begin{equation}
\label{emr0}
\begin{split}
E=-\epsilon_F+E_d+8\epsilon_F(k_Fa)/(3\pi)+O(a^2)
\end{split}
\end{equation}

\subsubsection{Full variational treatment}
In the general case $\left(\ref{molen}\right )$, up to one particle-hole pair, we can obtain a closed equation:
\begin{equation}
\begin{split}
&\frac{E+\epsilon_F-\epsilon_{p+q-k}/2-\epsilon_k+\epsilon_q-\nu}{g^2}\beta_{kq}=\\
&\frac{\frac{(k-p/2)(q+k-p)}{2E_k}\frac{1}{L^2}\sum_{k'q'}'\frac{(k'-p/2)(q'+k'-p)}{2E_{k'}}\beta_{k'q'}}{\frac{E+\epsilon_F-\epsilon_p/2-\nu}{g^2}-\sum_{k'}'\frac{(k'-q/2)^2}{LE_{k'}}}\\
&+\frac{(q+k-p)}{2L}\sum_{q'}'\frac{(q'+k-p)}{2E_k}\beta_{kq'}\\
	&+\frac{1}{L}{\sum_{k^\prime}}'\frac{(2k^\prime+k-p-q)^2}{4E_{k^\prime kq}}\beta_{kq}\\
	&-{\sum_{k^\prime}}'\frac{(2k^\prime+k-p-q)(k^\prime+2k-p-q)}{4LE_{k^\prime kq}}\beta_{k^\prime q}
\end{split}
\label{m}
\end{equation}
where
\begin{equation}
\begin{split}
E_k&=E+\epsilon_F-\epsilon_{p-k}-\epsilon_k\\
E_{k^\prime kq}&=E+\epsilon_F-\epsilon_{p+q-k-k^\prime}-\epsilon_k-\epsilon_{k^\prime}+\epsilon_q
\end{split}
\end{equation}
and Eq. \ref{m} can be formulated in a matrix form
\begin{equation}
\frac{1}{L^2}\sum_{k^\prime q^\prime}M(kq,k^\prime q^\prime)\beta_{k^\prime q^\prime}=0
\label{mol2}
\end{equation}
\begin{figure}[H]
  \centering
  \includegraphics[width=0.45\textwidth]{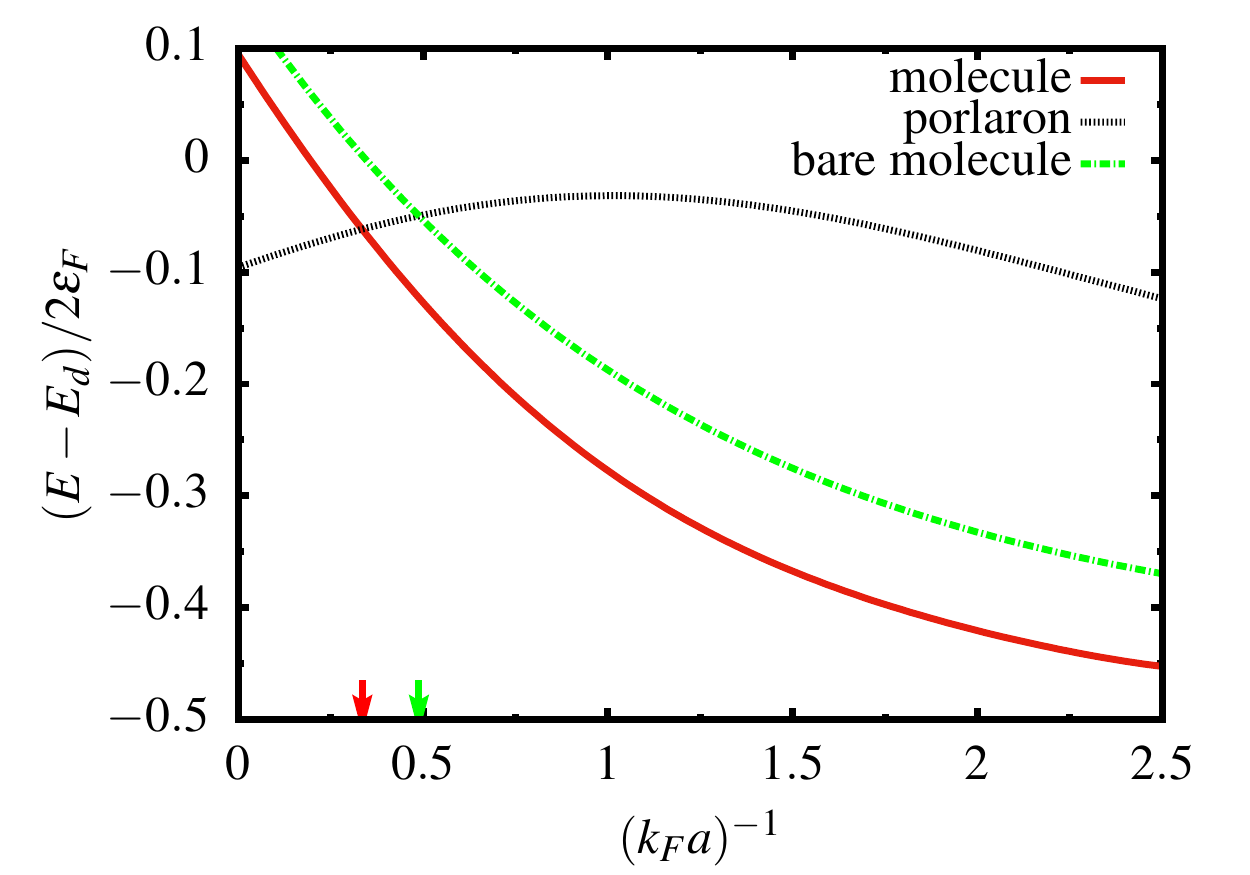}\\
  \caption{Ground-state energies for various states as a function of $(k_Fa)^{-1}$ with $k_F|r_0|=0.2$. The red solid line shows the molecule dressed with single-particle-hole excitation, the dashed line polaron with single-particle-hole excitation, green dot-dashed line the molecule without particle-hole excitation. The vertical left red and right green arrows locate the transition points from polaron to dressed molecules and bare molecules, respectively..}\label{diffpm}
\label{mpen}
\end{figure}
The ground-state energy $E$ is now simply obtained by the condition that the determinant of $M(kq,k^\prime q^\prime)$ vanishes. We evaluate the determinant numerically by discrediting the integral equation using a Gauss-Legendre quadrature and calculating the determinant of the corresponding linear equation system. The ground-state energy as a function of $(k_Fa)^{-1}$ is shown in Fig. \ref{mpen} where we plot the energies of the attractive polaron and molecule calculated with Eqs. \ref{polen2},\ref{nophm}, and \ref{mol2} at $p=0$. We can see that the polaron-to-molecule transition can occur as the interaction increases.

\begin{figure}[H]
  \centering
  \includegraphics[width=0.45\textwidth]{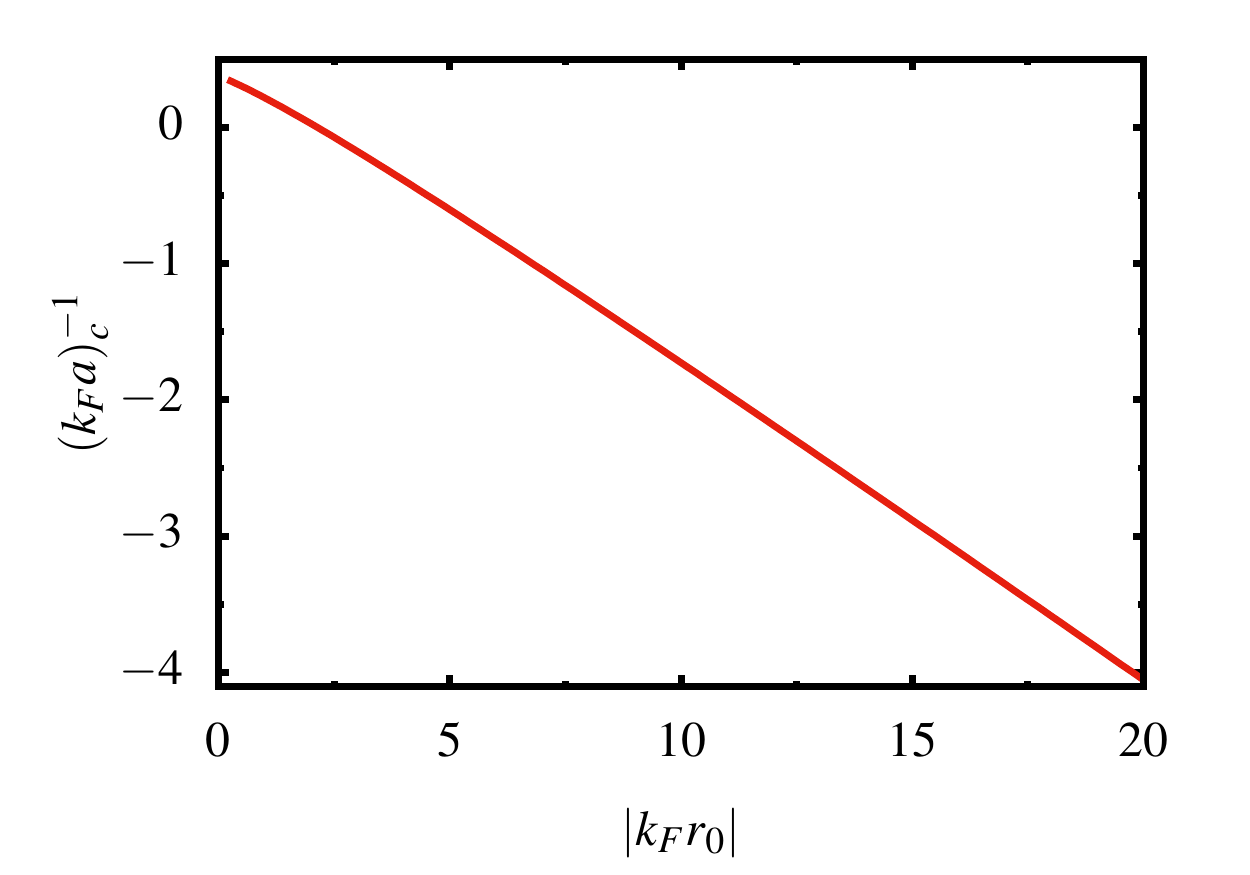}
  \caption{Critical interaction strength of the polaron-to-molecule transition as a function of the effective range. Above (below) the line the ground state is a molecule (polaron).}\label{kac}
\end{figure}
In Fig. \ref{kac} we plot the polaron-to-molecule transition point in the $(|k_Fr_0|,(k_Fa)^{-1})$ plane. We can see that as $|r_0|$ increases, the transfer point $(k_Fa)^{-1}$ moves to the BCS side of the $p$-wave resonance.

It is useful to compare the p-wave and s-wave polaronic systems in 1D. For one dimension $s$-wave, it is shown that there is no polaron-molecule transition \cite{McGuire1966,Giraud2009a,Parish2011,PhysRevA.94.043645}. As shown in \cite{Parish2011}, because of the divergent problem, the energy of $|\cal{M}\rangle$ is always higher than $|\cal{P}\rangle$ for all coupling constants, and therefore there are no poarlon-molecule transitions. This is different from the $p$-wave polaron, where we can find such a transition for both bare and dressed molecules. In fact, the 1D $p$-wave system has many similarities to 3D $s$-wave systems, as previously pointed out for the two-body scattering properties \cite{Cui2016}. Here we find that the 1D $p$-wave system also exhibits a polaron-molecule transition, similar to 3D $s$-wave system.

\section{CONCLUSION}
\label{conclusion}
In this paper we studied the 1D highly polarized $p$-wave Fermi gas across a narrow p-wave resonance. Using the variational approach up to single-particle-hole excitations, we have calculated the energy, residue, and effective mass of the polaron state, and compared the polaron energy to the energy of molecules, which shows the polaron-molecule transition as the interaction increases. We also studied the effect of a finite effective range to the polaron and molecule energies and their transition points. In particular, it was found that the critical interaction for the polaron-molecule transition shifts to the BCS side of the $p$-wave resonance as the effective range increases. Our results may be detected in quasi-1D Fermi gas across narrow $p$-wave resonances.

\section*{ACKNOWLEDGMENT}

The work is supported by the National Key Research and Development Program of China (2018YFA0307600, 2016YFA0300603), and the National Natural Science Foundation of China (No.11622436,No.11421092, No.11534014).

%\nocite{*}
\bibliographystyle{apsrev4-1}

\end{document}